# Polymer patterning by laser-induced multi-point initiation of frontal polymerization

Andrés L. Cook[1], Mason A. Dearborn[2], Trevor M. Anderberg[1], Kavya D. Vaidya[2], Justin E. Jureller[3], Aaron P. Esser-Kahn[2*], Allison H. Squires[2*]

[1]Department of Physics, University of Chicago
[2]Pritzker School of Molecular Engineering, University of Chicago
[3]James Franck Institute, University of Chicago
* To whom correspondence should be addressed: aesserkahn@uchicago.edu and asquires@uchicago.edu

**Abstract**

Frontal polymerization (FP) is an approach for thermosetting plastics at lower energy cost than an autoclave. The potential to generate simultaneous propagation of multiple polymerization fronts has been discussed as an exciting possibility. However, FP initiated at more than two points simultaneously has not been demonstrated. Multi-point initiation could enable both large scale material fabrication and unique pattern generation. Here the authors present laser-patterned photothermal heating as a method for simultaneous initiation of FP at multiple locations in a 2-D sample. Carbon black particles are mixed into liquid resin (dicyclopentadiene) to enhance absorption of light from a Ti:Sapphire laser (800 nm) focused on a sample. The laser is time-shared by rapid steering among initiation points, generating polymerization using up to seven simultaneous points of initiation. This process results in the formation of both symmetric and asymmetric seam patterns resulting from the collision of fronts. The authors also present and validate a theoretical framework for predicting the seam patterns formed by front collisions. This framework allows the design of novel patterns via an inverse solution for determining the initiation points required to form a desired pattern. Future applications of this approach could enable rapid, energy-efficient manufacturing of novel composite-like patterned materials.

**Introduction**

Frontal polymerization (FP)[1,2] is an energy-efficient alternative manufacturing mode for thermoset polymers and composites, which require large amounts of energy for thermal processing in a pressurized autoclave.[3] In contrast, FP is a self-propagating polymerization reaction that requires only a small, localized energy input for initialization, typically delivered by a heated tip or wire.[2,4,5] After initiation, heat is generated by the exothermic polymerization reaction and propagated by conduction and convection.[4,6] FP was initially described using acrylate chemistry, but has since been demonstrated using many other monomers, although largely within the acrylate family.[4] In this work, we focus on polymerizing dicyclopentadiene (DCPD), in which FP proceeds by frontal ring-opening metathesis polymerization (FROMP).[7–9]

Beyond its potential advantages for energy efficiency, FP can also be used to produce patterned polymers with spatially varying material properties.[10–17] To date, the best-characterized means of FP pattern creation have used single fronts to create nonuniformity, multiple materials to create a copolymer or composite,[11,12,15] or modulation of the front (or material[18]) to pattern a single polymer.[10,13,14,17,19] Less explored is the use of multiple fronts in patterning, which prior review work has identified as a target for future development of FP technology.[4] When two fronts meet, they form a seam at the collision zone.[20] Because seams polymerize at a higher temperature than the surrounding polymer,[21,22] controlled seam formation allows for controlled energy deposition. More colliding fronts would allow more complex patterning. However, the position of a seam depends on the difference between the initiation times of the colliding fronts. Multipoint patterning is therefore limited by number of near-uniform initiation times, which becomes increasingly difficult with more initiation points and spatial resolution.

Photoinitiation is a promising route to overcoming challenges in implementing multipoint initiation because light can be readily patterned onto a substrate to achieve concurrent rather than sequential initiation at multiple locations. To date, light has been used to trigger FP either using a photoinitiator that chemically or thermally triggers polymerization,[23–28] or using direct absorption of light by the sample.[29–32] Lasers offer high levels of power per unit area in a slowly diverging beam, which has enabled demonstration of photoinitiation of FP from a distance.[30–33] In principle, lasers can also offer precise spatiotemporal control of the beam by steering from a distance with galvanometer mirrors or acousto-optic deflectors. A steered laser beam could be programmed to pause at each of several desired initiation locations arranged in any geometric pattern, enabling highly reproducible photoinitiation patterning. This approach contrasts with patterned FP by resistive heating, where each wire operates independently and must be mechanically placed at the initiation location.

Here we accomplish multi-point initiation in DCPD for up to 7 points using a focused, steered laser beam time-shared among initiation points. We employ carbon black to assist photothermal initiation via increased heat absorbance of patterned near-IR laser illumination. To enable optimization of laser-patterned multi-point FROMP initiation, we first characterized front initiation and propagation parameters for varying optical and geometric conditions. Features of the composite resulting from front intersections, including seams and peaks, are governed by the geometry of the initiation pattern. We developed a geometric framework for understanding and predicting these features for any given pattern of initiation points. We apply this framework to the inverse problem to solve for the initiation point pattern that will produce a desired set of seams. Finally, we demonstrate that our initiation mechanism also allows us to use advanced initiation methods including asynchronous and linear initiation patterns. This work is the first instance of simultaneous FP initiated at more than two points and presents new opportunities for FP-based polymer patterning and manufacturing.

**Results**

**Programmable laser-ignited frontal polymerization**
In developing a method to create complex and precise patterns of seams in DCPD, we used a pulsed Ti:Sapphire laser tuned to 800 nm to heat the monomer, which is doped with carbon black to increase light absorption (Figure 1a). We coupled this laser to a pair of galvanometer mirrors to allow us to rapidly change the position of the beam (Figure 1b). By steering the laser focus quickly among multiple initiation points with dwell times of 1-5 ms each, we can effectively time-share the beam power, as shown in Figure 1c. This results in fast, synchronized multipoint initiation of FP, demonstrated in the thermal snapshots of Figure 1d. After initiation, the fronts move outward radially until they collide, polymerizing the entire sample. As expected, the seams where fronts collide reach higher maximum temperatures than the surrounding polymer, indicating that seam patterning corresponds to patterning of thermal energy within the material.

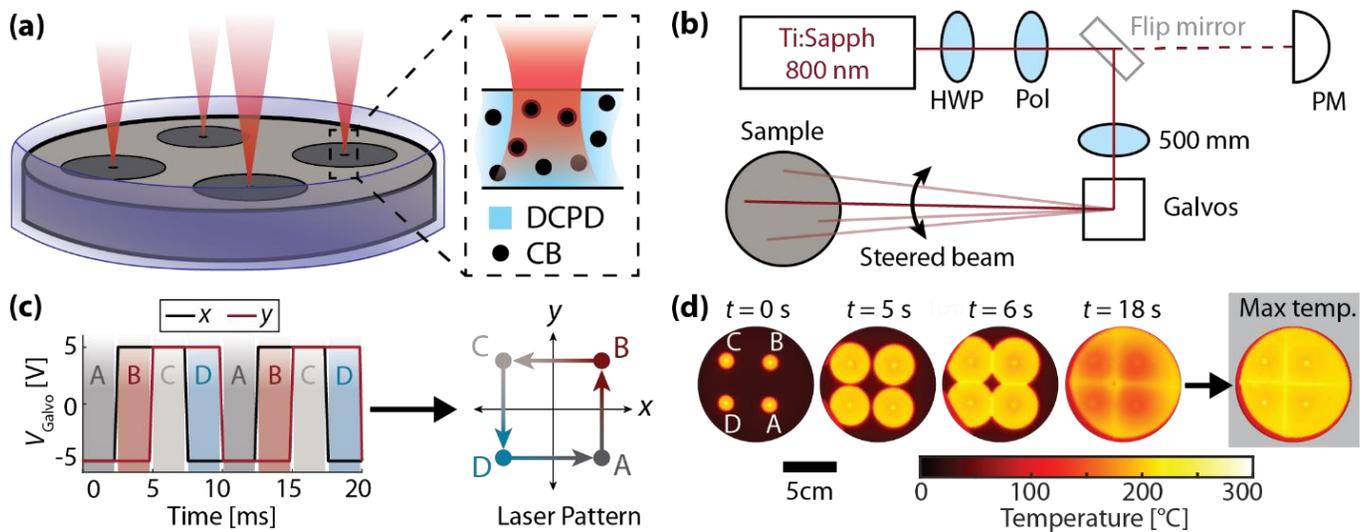

Figure 1: Overview of multipoint frontal polymerization. **(a)** Illuminating carbon black-doped DCPD from above with a near-IR laser causes rapid heating, followed by initiation of FP. **(b)** Schematic of optical apparatus for steering the focused laser beam on the sample. The beam first passes through a half-wave plate (HWP) and polarizer (Pol), decreasing its power according to the wave plate angle. It can then be directed either into a power meter or through a lens (focal length 500 mm) and into a pair of movable galvanometer mirrors (galvos). The laser is then directed onto the sample by an overhead mirror. **(c)** Illustration of beam time-sharing by galvos. Left: Input voltage signals as a function of time for *x* and *y* galvo mirrors. Right: Path of the beam is a parametric curve in the *x-y* plane, as created by the time-dependent voltage signals at left. **(d)** Frames from a thermal video of a four-point polymerization, showing fronts propagating and colliding with one another. The far-right panel shows the maximum temperature of each pixel over time, showing that seams polymerize at a higher temperature than the bulk.

## Characterizing Laser-Ignited FP

To quantify the rapidity and synchronicity of initiation, we measured initiation onset time and time between initiation points.[28,29]. Figure 2a shows the time to FROMP initiation as a function of the applied laser power for a series of single-point tests. As the power increases, the initiation time converges towards zero. These data are well-fit by a simple model where initiation occurs once a fixed amount of energy $E$ has been delivered to the sample, so that the time to initiation $t$ is inversely proportional to the initiation power $P$:

$$t = \frac{E}{P} \tag{1}$$

The constant of proportionality is the energy required to initiate FP. Here, a weighted fit to our data gives a value of *P* = 294 ± 108 mJ. For more details regarding the trend shown in Figure 2a, see SI Note 1.

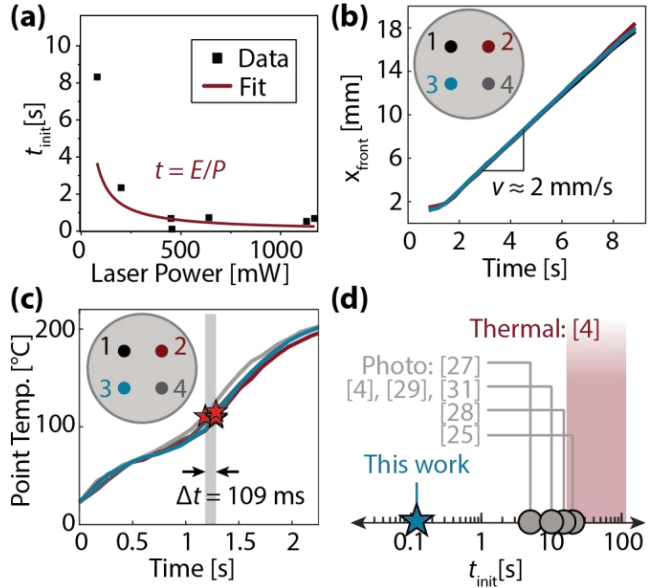

Figure 2: Characterization of multipoint FP. **(a)** Time to initiation vs. illumination power, along with the reciprocal fit curve. The fit indicates a fixed energy for initiation of 294 mJ. **(b)** Front position as a function of time for four simultaneously ignited fronts, showing a constant front velocity of 2 mm s$^{-1}$. **(c)** Initiation point temperature vs. time for a four-point initiation. Red stars indicate the identified initiation times. **(d)** Photoinitiation times from literature (gray) and our work (blue). Where one work gave multiple initiation times, the shortest was used.

The required synchronicity for patterning depends upon the front propagation speed and the desired placement precision of features such as seams. In our system, the front propagation speeds are nearly identical within each sample at approximately 2 mm s$^{-1}$ (Figure 2b). Therefore, reproducible feature placement requires synchronicity of FP initiation within ~1 second (see SI Note 2). The measured synchronicity of our setup is of order tenths of a second. To demonstrate this, Figure 2c shows temperature traces at each of four initiation points during illumination. An initial sharp temperature rise plateaus slowly, followed by FP initiation at about 1.25 s (gray band), with exact initiation times calculated as described in SI Note 1. Here, all points initiate within 0.11 s of each other. For further discussion of the causes of asynchronicity, see SI Note 2. Beyond its advantages in speed, laser-ignited multipoint FP is also highly synchronous.

Rapid initiation is helpful for patterning, as it increases synchronicity and reduces diffusive pre-heating of the monomer surrounding the initiation points. In comparing our method with the prior literature, we noted that our laser-based method initiated FROMP much more rapidly than other light-based setups. Representative initiation times from previous photoinitiation studies are shown in Figure 2d. Our approach decreases initiation time by over an order of magnitude while retaining the ability to initiate at many individual, arbitrarily patterned points.

**Controllable seam patterning with multipoint FP**

To understand how multipoint FP generates geometric patterns, we developed a model based on the geometry of the initiation points. Two example geometries that we tested are shown in Figure 3a, one with three points (top) and one with seven points (bottom). Under the assumption that polymerization fronts propagate radially outward at constant and equal speeds, as was observed in our original tests (Figure 2b), we can predict where seams will form. When polymerization is initiated simultaneously at multiple points, pairs of fronts will first intersect at the midpoint between their corresponding initiation points. As the fronts continue to propagate, they meet at points equidistant from each initiation point: the perpendicular bisector of the initiation points. When more than two initiation points are present, the perpendicular bisectors form a network called a Voronoi diagram. Therefore, using the initiation point geometries from Figure 3a, we predicted that seams would form as shown in Figure 3b.

Having developed our model, we then verified it experimentally. We implemented both the three and seven-point patterns from Figure 3a with initiation patterns as shown in the thermal images in Figure 3c. Images of the resulting polymerized samples are shown in Figure 3d, illustrating that the seam patterns match the predicted locations. In addition to seams, we noticed the formation of peaks in the sample at locations where three fronts collided, as well as some minor rippling features, both of which are evident from contrast changes in the digital image. Figure 3e shows the measured locations of all initiation points, seams, and peaks superimposed on the image. These results can be extrapolated to the more general observation that mathematically, any set of initiation points result in a seam pattern that can be modeled and predicted using a Voronoi diagram.

The inverse problem is to derive the initiation point locations required to produce a desired seam pattern, such as the one shown in Figure 3f. Using a geometric algorithm for inverse Voronoi diagrams[34], we predicted that this seam pattern could be produced by the set of initiation points shown in Figure 3g. We then implemented this pattern as shown in the thermal image in Figure 3h, which resulted in the correct polymerized sample pattern shown in Figure 3i. Measured locations of all initiation points, seams, and peaks are superimposed onto the digital image in Figure 3j. In contrast to the forward prediction problem, not all desired seam patterns can be realized by FP because not all sets of connected line segments form Voronoi diagrams. Moreover, some desired seam patterns have degenerate inverse solutions (See SI Note 2).

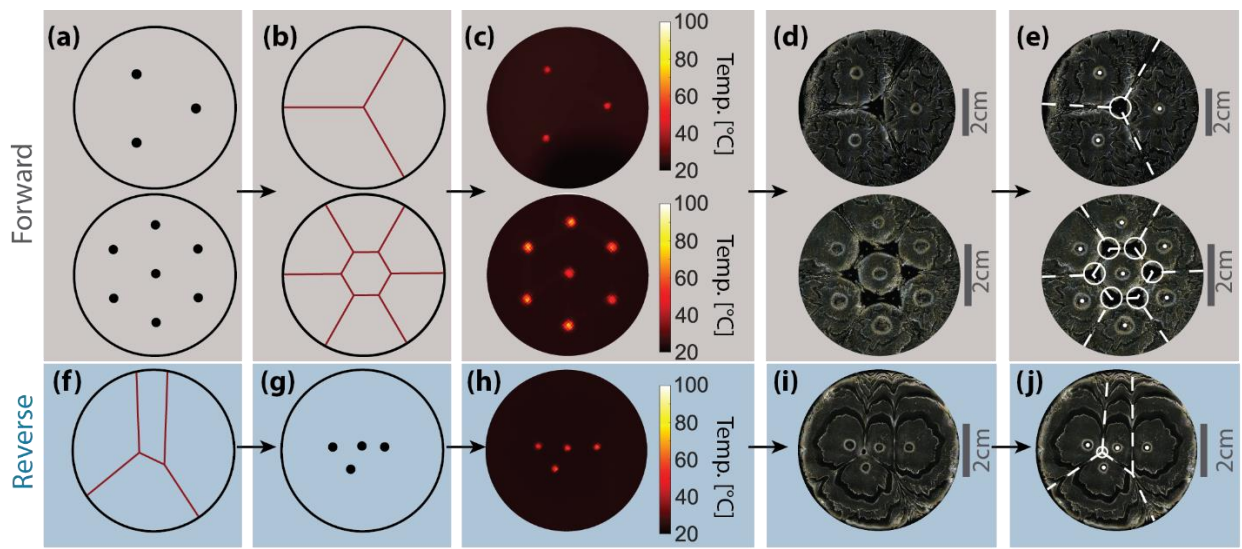

Figure 3: Process for forming multipoint FP pattern. **(a)** Desired initiation pattern. **(b)** Predicted seam pattern. **(c)** IR image of initiation points. **(d)** Camera image of FP result. **(e)** Camera image, annotated with seams (dashed lines), initiation points (closed circles), and peaks (open circles). Panels **(f)**-**(j)** show the inverse Voronoi algorithm to produce an initiation pattern from a desired seam pattern. The first two panels are therefore reversed from **(a)** and **(b)**.

**Controllable peak patterning with multipoint FP**

As seams form, they can collide with each other to form new structures at their intersection, as depicted in Figure 4a. Because fronts propagate at equal speeds, this intersection will be located at the point equidistant from all initiation points involved, i.e. their circumcenter. If the circumcenter is inside of the polygon formed by the initiation points, then the expanding fronts will enclose a "trapped volume" before they all converge at the center (top panels of Figure 4a). As the trapped volume contracts, we hypothesize that unreacted monomer is pushed ahead of the front and into the center, forming a peak at the intersection. We expect to only see peaks forming when the initiation points form a polygon that contains its circumcenter. Otherwise, two seams will collide and form a third seam, with no peak present, as depicted in the bottom panels of Figure 4a. Our success in predicting and patterning seam locations also allowed us to predict and pattern the locations of elements that are formed by the intersections of seams.

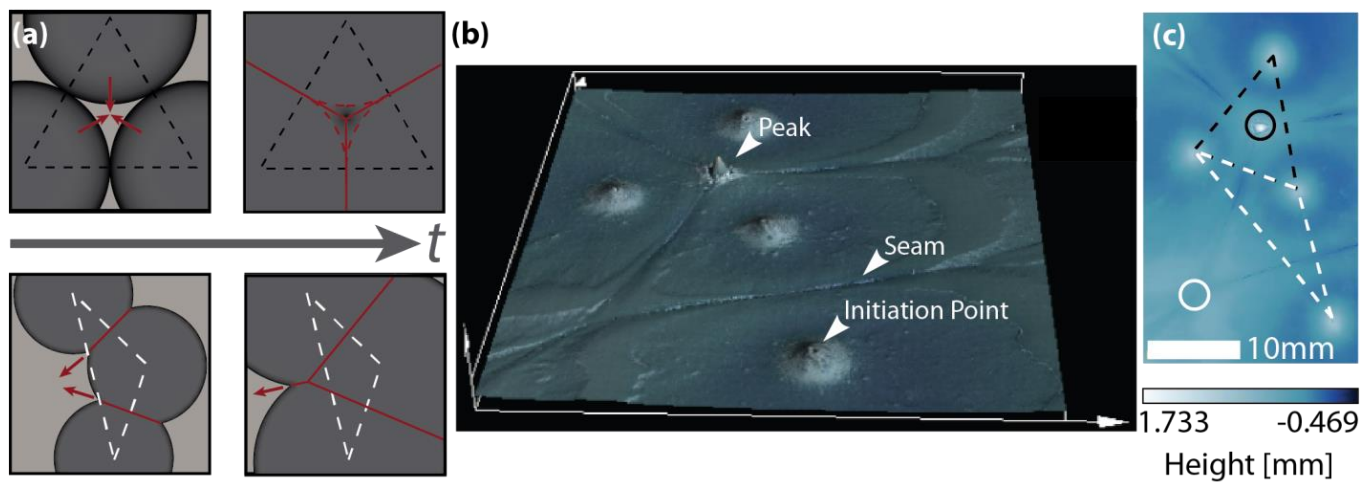

Figure 4: Characterization of surface patterns induced by multipoint FP. **(a)** Cartoon schematic of peak formation, showing seams forming (red) and monomer pushed away from fronts (red arrows). Dashed triangles connect the initiation points involved in seam convergence. **(b)** 3D scan of a sample showing two seam convergences, one with and one without a peak (false color overlay indicates height) (not tilt corrected). Image region is 36 mm wide (horizontal page axis) by 39 mm high (vertical page axis). **(c)** Heightmap of the sample from **(b)**. An acute triangle (black) results in a peak when the seams intersect inside the triangle, while an obtuse triangle (white) does not lead to a peak.

To demonstrate the principles of peak formation, we examined a sample that shows both peaked and non-peaked seam collisions. A 3-D profile of this sample shows major surface features including seams, peaks, and initiation points (Figure 4b). The four initiation points shown form two triangles, one acute and one obtuse (Figure 4c). Because acute triangles contain their circumcenters while obtuse triangles do not, we expect to see a peak at the acute circumcenter and no peak at the obtuse circumcenter, as borne out by our experimental results and illustrated in the height map (Figure 4c, circles). Based on our model, only patterns with straight seams are achievable with simultaneous multi-point FP initiation, and the geometry of initiation points is dictated by formation of seams at midpoints between them.

**Asynchronous and non-punctate FP initiation patterns**

Asynchronous multipoint initiation could provide access to more complex patterns. While this technique pushes the limits of our current abilities, we can demonstrate a simple example experimentally. When we initiate polymerization at two points at different times, the fronts will no longer meet along a straight line. Instead, the seam will form at points where the distance to one initiation point is equal to the distance to the other initiation point, plus a fixed offset $v_f \Delta t$, where $v_f$ is the front speed and $\Delta t$ is the delay in initiation times. This condition defines a hyperbola bending around the point that initiates later. To demonstrate this effect, we performed an asynchronous initiation of two points by increasing the time per galvo cycle spent on one initiation point relative to another, as shown in Figure 5a. As expected, the resulting seam forms a hyperbola.

Similarly, non-punctate initiation patterns also have the potential to generate complex seam and peak patterns. To generate a line or curve of initiation rather than a point, we smoothly scan the galvos by feeding each motor a (sampled) continuous function, uniformly heating a contour. We implemented this for a circular path, producing both an inward-propagating front that forms a peak at the ring's center, as well as an outward-propagating front. The resulting polymer sample is shown in Figure 5b; it is unclear why the ring is distorted from the circle projected by the laser. Together, the ability to apply non-punctate and asynchronous patterns to initiate FP substantially expands the range of complexity that can be patterned in the resulting material.

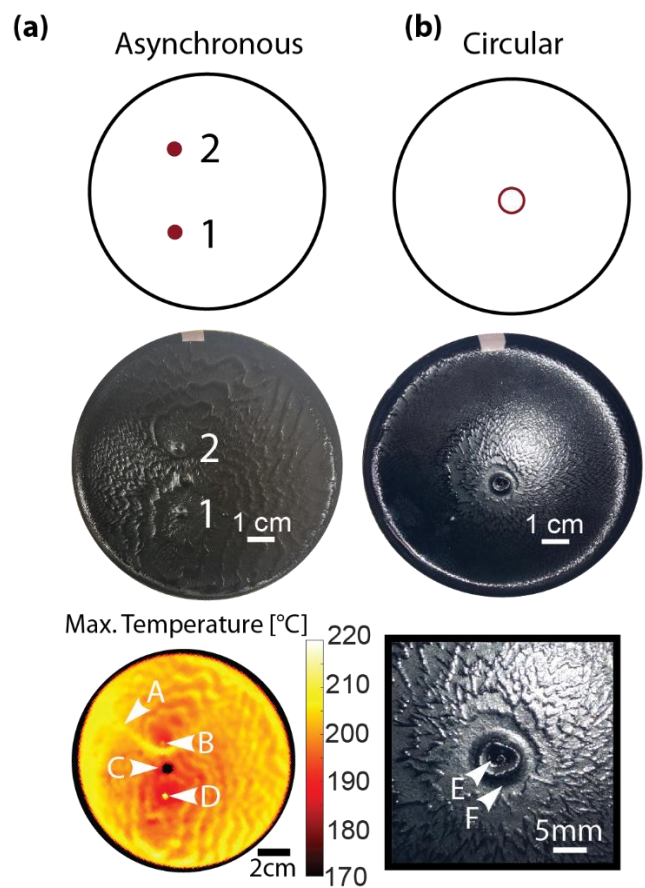

**Figure 5:** Asynchronous and circular initiation. **(a)** Asynchronous initiation at points 1 and 2 (left column). **(b)** Circular initiation (right column). Top row: initiation point patterns. Middle row: camera images of resulting sample. Bottom row: Maximum temperature. Labeled points: A: Hyperbolic seam, B: second initiation point, C: carbon black cluster, D: first initiation point, E: inner peak, F: initiation ring.

**Discussion and Conclusion**

Multipoint FP has great potential for producing patterned materials but has previously been limited by poor synchronicity. Here, we demonstrate a new approach for photothermal initiation by patterning a focused laser beam on a sample, allowing us to initiate polymerization simultaneously at up to 7 points and much more rapidly than previously reported. Aside from its speed, our laser-ignited FP is well-controlled, with synchronous initiation and steady front speeds. Because of this controllability, we can describe the behavior of the fronts with simple rules that allow us to design new patterns. These designs can be implemented rapidly due to the flexibility of time-shared laser ignition, providing a fast iterative design loop for patterned polymer manufacturing.

While we can initiate multipoint FP with unprecedented precision, our approach is not without limitations. Synchronicity is limited by the stochastic nature of initiation (see SI Figure 2). This is potentially due to poor sample mixing, flow conditions during heating, or the presence of carbon black clusters within the sample. The carbon black itself also limits the optical properties of the polymers and increases the viscosity of the monomer mix. Additionally, the time-sharing system used to split the beam limits the number of points that can be accessed before the laser's power is spread too thin to initiate quickly. Consistent with other FP studies, different batches of monomer and catalyst often result in different surface artifacts (such as ripples and pulses).

Future avenues of multipoint FP research should focus on either characterizing existing multipoint processes or expanding the space of constructible patterns. While our initiation time study (Figure 2b) was conducted with single-point initiation, a similar study on multipoint initiation could provide more insight into the underlying energetics. Similarly, an investigation of the initiation point temperature profiles may explain the anomalies described in SI Note 2. To expand our existing space of patterns, future studies can better define patterning rules and inverse algorithms for asynchronous and non-punctate initiation. Additionally, our framework currently assumes a planar sample, but the geometry would become substantially more complex on a curved surface. A differential-geometric approach would allow new patterns to be created on non-flat surfaces.

Multipoint FP has broad applications in polymer manufacturing. Seam patterning might be used directly to create inhomogeneous materials with composite properties, despite having uniform chemical composition. We can also exploit the uneven distribution of thermal energy to power secondary processes, from chemical reactions to the deformation of heat-memory alloys. Furthermore, laser-ignited multipoint initiation can be used outside of patterning to polymerize objects too large for a single front, or the heat deposition can be altered dynamically as part of a front-control feedback loop, such as that designed by Schaer and Bretl[35]. Multipoint initiation, previously inaccessible, is now opening up new opportunities in component manufacturing and design, expanding the potential of FP-based methods.

**Methods**
**Sample preparation**
The monomer preparation procedure is derived from Robertson et al.[2] 5-ethylidene-2-norbornene (ENB) is added as a stabilizing agent to prevent the DCPD from solidifying at room temperature. We added 625.7 µL phenylcyclohexane (PCH) to 256.8 µg second-generation Grubbs catalyst (GC2) and sonicated for 20 min to dissolve, after which 37.2 µL tributyl phosphite (TBP) was added as an inhibitor and mixed by inversion. The combined catalyst and inhibitor was dispensed into 626 µL aliquots and stored at -4 °C prior to use.

Prior to polymerization, 950 mg of carbon black was added to 20 mL of stabilized monomer and sonicated for 20 min, with manual stirring after 10 and 20 minutes of sonication. The sample was then stored short-term in a rotary mixer to keep the carbon black suspended until polymerization (no longer than 2 h). Immediately before polymerization, the thawed catalyst/inhibitor was added to the monomer, mixed by inversions, and poured into a 10 cm diameter petri dish (Corning 3160102BO). The thickness of the poured sample prior to polymerization is approximately 2.5 mm.

**Laser apparatus**

The optical setup for photothermal initiation depicted in Figure 1b shows near-IR laser illumination passing through power and position control optics to bring it to the sample plane: The light source is a Coherent Chameleon Ultra Titanium:sapphire 800 nm pulsed laser. Immediately prior to polymerization (after combining the monomer and catalyst), we read the laser power with a power meter placed on an alternate beam path, switchable by a flip mirror. The laser passes through a 500 mm focal length lens that focuses the beam to a 0.255 mm-wide point on the surface of the sample.

To control the position of the beam, we send the laser through a Thorlabs QS15XY dual galvanometer mirror, one controlling each axis of motion, controlled by an Arduino microcontroller. The galvos are driven by a pair of -5 V to 5 V analog electrical signals (one per axis), while the Arduino outputs a pair of digital signals representing two numbers between 0 and 4095, inclusive. To mediate between these two signals, we use a custom "predriver" board composed of a digital-analog converter (DAC) and amplifier for each axis. The DAC converts the digital signal to an analog signal between 0-5 V, which the amplifier converts to the -5 V – 5 V range required by the galvos. Through these electronics, we can represent points on the sample as a pair of integers, each from 0 to 4095. Patterns are programmed into the Arduino as three separate lists. The first two lists store the *x* and *y* coordinates of each point in the sequence, while the third list stores the dwell time. The Arduino then iterates through the lists, holding on each (*x*, *y*) position for the specified time. When the end of the list is reached, the control software loops back to the beginning. Because the galvo controls the angle of the beam rather than its position (in the plane transverse to the beam), scaling of the galvo signal to *x-y* positions on the sample depends upon the distance between the galvos and the sample.

**Temperature measurement**
Temperatures were recorded with a FLIR E8 infrared camera mounted above the sample. For each video, frames were exported from FLIR ResearchIR to a stack of .csv files (one per frame) containing temperature values (16-bit floating point) that were imported into MATLAB for analysis.

**Height profiling**
Height profiles of polymerized samples were taken on an Olympus DSX1000 microscope. Volume estimates were made using the LEXT OLS5100 analysis application.

**Analysis:**
*Image scaling:* To determine the appropriate pixel-to-millimeter scale factor, the diameter of the dish (10 cm) was used as a reference. Because images are typically taken from above, we do not expect obliquity to affect the scaling. In cases where the Olympus microscope was used (i.e. height maps and Figure 3e and 3j), the instrument provided a scale bar.

*Front speed:* To determine the speed of the front, we first determined its displacement by measuring the area enclosed by a front and converting to the equivalent radius of a circle $\left(r = \sqrt{\frac{A}{\pi}}\right)$. This approximation has the effect of averaging over directions and limiting effects of anisotropic front propagation. Differentiating the front radius with respect to time results in the front velocity.

**Acknowledgements:**

The authors acknowledge valuable discussions and feedback from: Kepler Domurat-Sousa, Katie Kloska, Samantha Livermore, Yixiao Dong, Farsa Ram, and Sarah Brown. A.H.S., A.E.-K., and J.J. acknowledge support from NSF QLCI QuBBE grant OMA-2121044. A.E.-K. acknowledges support from AFOSR FA9550-18-1-0229 and FA9550-20-1-0194. A.H.S. acknowledges support from the Neubauer Family Foundation. This work made use of the shared facilities at the University of Chicago Materials Research Science and Engineering Center, supported by National Science Foundation under award number DMR-2011854.

**Data Availability Statement**
Raw and figure data are available from the corresponding authors upon reasonable request.

**Conflicts of Interest:**
The authors declare no conflict of interest.

**Table of contents text**

Frontal polymerization is a technique for rapid, energy-efficient polymer manufacturing in which polymerization proceeds outward from a point of initial external heating. When multiple fronts collide, the collision zone polymerizes at a much higher temperature than the bulk monomer, creating a seam. By simultaneously heating multiple points using a laser, the authors collide fronts to create programmably patterned materials.

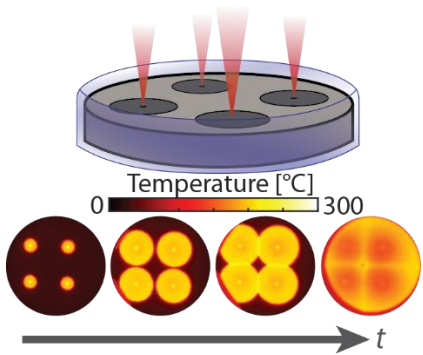

1
2
3
4
5
6
7
8

# SUPPORTING INFORMATION:
# Polymer patterning by laser-induced multi-point initiation of frontal polymerization

Andrés Cook[1], Mason Dearborn[2], Trevor Anderberg[1], Kavya Vaidya[2], Justin Jureller[3], Aaron Esser-Kahn[2*], Allison H. Squires[2,4*]

[1]Department of Physics, University of Chicago, Chicago, IL, USA
[2]Pritzker School of Molecular Engineering, University of Chicago, Chicago, IL, USA
[3]Materials Research Science and Engineering Center, University of Chicago, Chicago, IL, USA
[4]Institute for Biophysical Dynamics, University of Chicago, Chicago, IL, USA
* To whom correspondence should be addressed: aesserkahn@uchicago.edu and asquires@uchicago.edu

24  **Contents**

32

33

## SI Note 1: Initiation time calculation

In order to determine the time to initiation, we measure the temperature of individual initiation points. The temperature profile initially follows an exponential profile consistent with Newtonian heating. We define the time of divergence from this profile as the initiation time. To measure this, we specify a linear region of the Newtonian curve, fit a line to that portion of the temperature profile, and determine the first point at which the profile is at least 10% higher than the linear prediction. The resulting annotated temperature profiles are shown in SI Figure 1. The three vertical lines show the start and end of the fitting window (black), as well as the laser activation time (red), set as the common "zero time" for all profiles. The purple line is the linear fit, and the red X marks the initiation time.

The resulting power vs. time graph follows a reciprocal trend. This suggests a fixed energy cost to initiate polymerization:

$$t = \frac{E}{P} \tag{S1}$$

Where $E$ is the polymerization energy, $t$ is the initiation time, and $P$ is the initiation power. We determine $E$ by fitting Equation S1 to our experimental data, as shown in Figure **2a**. Because transport effects increase the variation in initiation time for large $t$, we weight our fit by the inverse of the initiation time. This fit results in a polymerization energy of approximately 300 mJ (294 mJ ± 108 mJ).

**SI Note 2: Pattern invertibility**

While we can generate initiation patterns for many desired seam patterns, not every seam pattern will give a unique initiation pattern. Some seam patterns can be created by infinitely many initiation patterns, some can be created by exactly one, and some are not invertible at all. SI Figure 3 shows examples of the first and last of those three cases. The seam pattern shown in SI Figure 3a has a continuum of possible initiation patterns: any rescaled version of the triangular pattern shown in blue will lead to the same seam pattern. By contrast, the seam pattern in SI Figure 3b has no inverse solutions. This is because any pair of initiation points that form a seam are reflections of each other about the axis of the seam. In order to form the 180º angle shown in the pattern, the leftmost initiation point must be a reflection of both points on the right, which is impossible. For more discussion of the mathematical underpinnings of inverse seam prediction, please see Ref. 34.

**SI Note 3: Synchronicity of initiation**

Shown in SI Figure 3 is a set of temperature traces for a four-point sample that displayed unusual lack of synchronicity, likely due to poor mixing or the presence of large clusters of carbon black particles. This asynchronicity, however, allows us to explore the factors limiting the efficacy of multipoint initiation. The temperature traces follow the exponential profile associated with Newtonian heating up until polymerization is initiated, at which point the temperature rapidly diverges. This divergence does not seem to occur at a fixed initiation temperature, but rather spontaneously at some point along the Newtonian heating profile. The apparent spontaneity of the divergence limits how simultaneously we can initiate polymerization, despite the laser's precision in delivering energy.

76 　　　　The traces of SI Figure 3 display a second unusual feature also found in other
77 temperature profiles: a "shoulder" occurring between 50 and 100 degrees Celsius. In this case,
78 the shoulder lasts an average of 1.25 s, while the fronts from that sample travel at an average
79 speed of 1.82 mm s$^{-1}$. These together indicate that the front travels 2.28 mm during the
80 shoulder, approximately equal to the 2.5 mm sample depth. This suggests that the shoulder is
81 caused by the fronts (which initiate at the top of the monomer due to carbon black's high
82 absorption) traveling to the bottom of the dish as a hemispherical wave before transitioning to a
83 cylindrical wave in the bulk monomer.

84 　　　　While we have not achieved perfectly synchronous initiation, we are able to impose
85 tolerances on the degree of synchronicity required to preserve a particular pattern. Because our
86 fronts travel with a speed on the order of mm s$^{-1}$ and have a width of order mm, the error in
87 front displacement should be of the same order of magnitude as the front width if different
88 points initiate within a few seconds of each other:

89 $$\Delta t \approx \frac{\Delta x}{v_f} \tag{S2}$$

90 On these grounds, we describe our initiation as synchronous.

91
92
93
94
95
96
97

**Supporting Figures:**

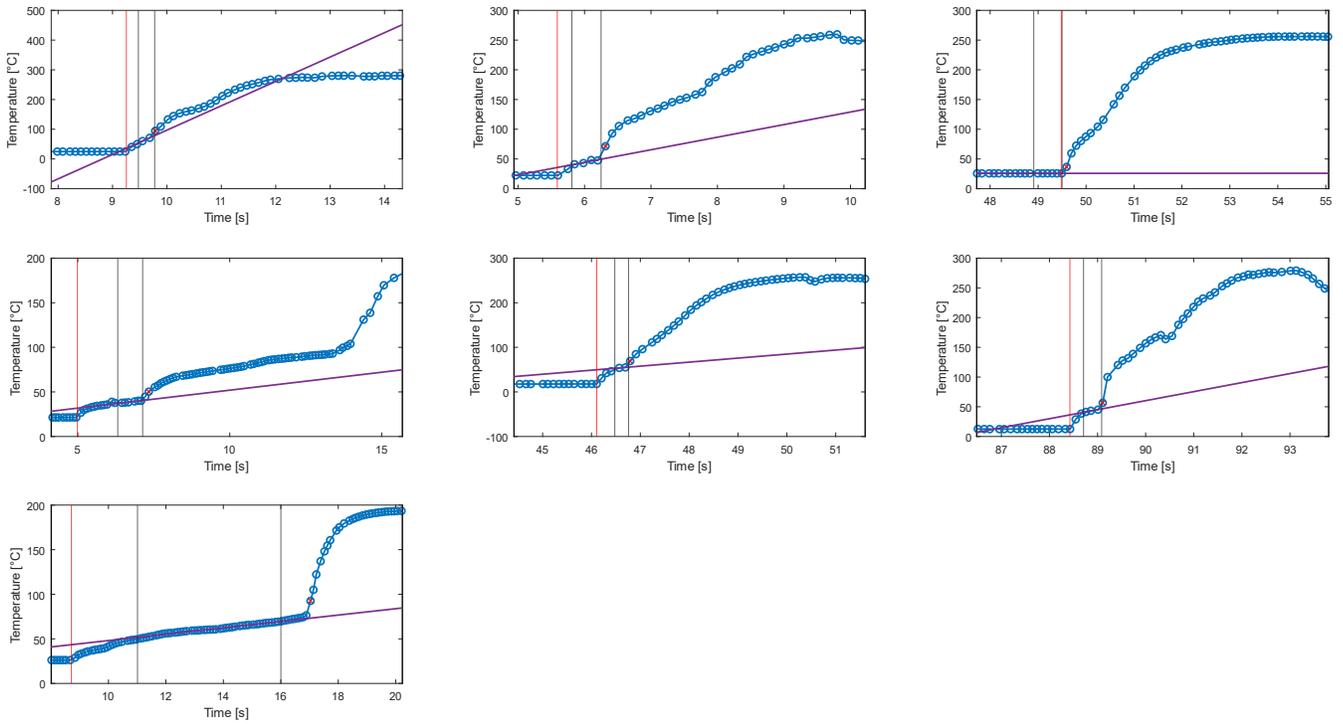

SI Figure 1: Example of initiation time calculation for seven FP samples, showing guidelines used to designate the fitting window (black) and laser activation time (red). The initiation time calculation algorithm produces a linear fit (purple) and initiation time (red x)

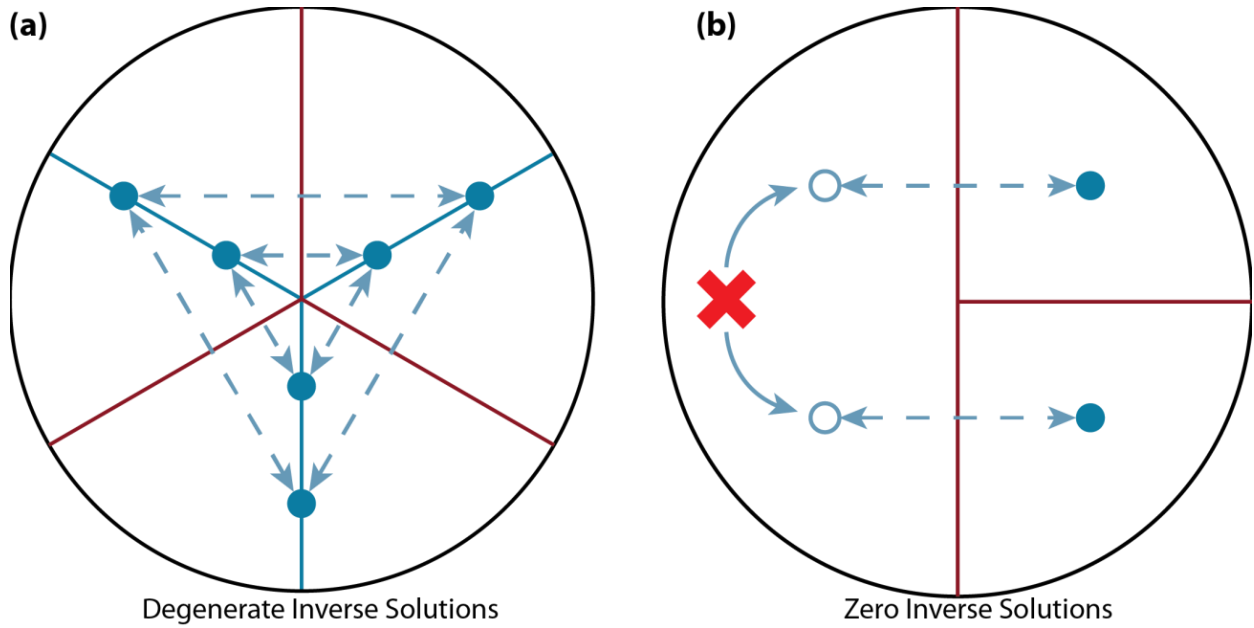

SI Figure 2: Seam invertibility. **(a)** A pattern with infinitely many inverses. Red lines show desired seams, while blue lines show possible initiation sites. Two possible sets of initiation sites, with seam bisectors, are shown in blue. **(b)** A pattern with no inverse. Again, red lines show desired seams and blue dots show initiation points. The required points to form the two "halves" of the vertical seam are shown as empty blue circles and are incompatible.

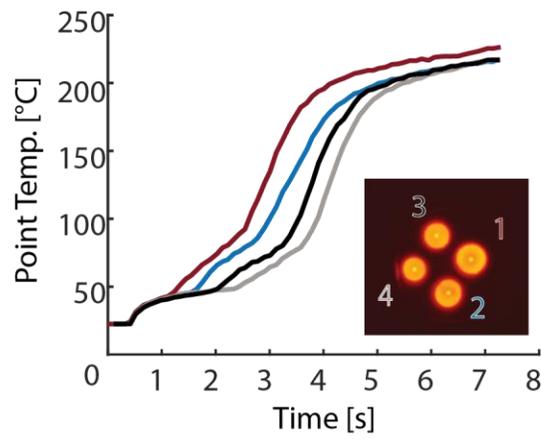

SI Figure 3: Temperature vs. time plot for a four-point sample. Inset: Thermal images with initiation points color-coded corresponding with the appropriate temperature curves.